\documentclass[aps,prl,preprint,superscriptaddress,floatfix,showpacs,nobibnotes]{revtex4-1}
\usepackage{graphicx,caption}
\usepackage{amssymb}
\usepackage{amsmath} 
\usepackage{amsfonts}
\usepackage{hyperref}
\usepackage{textcomp}
\usepackage{subfig} 
\usepackage{stackengine}
\usepackage{dashrule}
\usepackage{tabularx,comment,footnote}
\usepackage{float}
\usepackage{xtab}
\usepackage[style=base]{caption}
\usepackage{latexsym}
\renewcommand{\arraystretch}{1.5} 
\begin{document}

\title{Emergence of an island of extreme nuclear isomerism at high excitation near $^{208}$Pb}
\author{S.G. Wahid}
\email[Present address: ]{University of Massachusetts Lowell, USA}
\affiliation{School of Physical Sciences, UM-DAE Centre for Excellence in Basic Sciences, University of Mumbai, Mumbai 400098, India}
\author{S.K. Tandel} 
\email[Corresponding author: ]{sujit.tandel@cbs.ac.in ; sktandel@gmail.com}
\affiliation{School of Physical Sciences, UM-DAE Centre for Excellence in Basic Sciences, University of Mumbai, Mumbai 400098, India} 
\affiliation{Department of Physics, University of Massachusetts Lowell, Lowell, Massachusetts 01854, USA}
\author{Saket Suman}
\affiliation{School of Physical Sciences, UM-DAE Centre for Excellence in Basic Sciences, University of Mumbai, Mumbai 400098, India}
\author{P.C. Srivastava}
\affiliation{Department of Physics, Indian Institute of Technology Roorkee, Roorkee 247667, India}
\author{Anil Kumar}
\affiliation{Department of Physics, Indian Institute of Technology Roorkee, Roorkee 247667, India}
\author{P. Chowdhury} 
\affiliation{Department of Physics, University of Massachusetts Lowell, Lowell, Massachusetts 01854, USA}
\author{F.G. Kondev}
\affiliation{Argonne National Laboratory, Argonne, Illinois 60439, USA}
\author{R.V.F. Janssens}
\affiliation{Department of Physics and Astronomy, University of North Carolina at Chapel Hill, Chapel Hill, North Carolina 27599, USA}
\affiliation{Triangle Universities Nuclear Laboratory, Duke University, Durham, North Carolina 27708, USA}
\author{M.P. Carpenter} 
\affiliation{Argonne National Laboratory, Argonne, Illinois 60439, USA}
\author{T. Lauritsen}
\affiliation{Argonne National Laboratory, Argonne, Illinois 60439, USA}
\author{D. Seweryniak}
\affiliation{Argonne National Laboratory, Argonne, Illinois 60439, USA}
\author{S. Zhu}
\email[Deceased]{}
\affiliation{Argonne National Laboratory, Argonne, Illinois 60439, USA}

\date{\today}

\begin{abstract}
Metastable states with {\it T}$_{1/2}$ = 8(2) ms in $^{205}$Bi and 
{\it T}$_{1/2}$ = 0.22(2) ms in $^{204}$Pb, with $\approx $ 8 MeV 
excitation energy and angular momentum $\ge $ 22 $\hbar $, have been 
established. These represent, by up to two orders of magnitude, the 
longest-lived nuclear states above an excitation energy of 7 MeV, 
ever identified in the nuclear chart. Additionally, the half-life
of the 10.17 MeV state in $^{206}$Bi has been determined to be
0.027(2) ms, the next highest value in this highly excited regime.
These observations indicate the emergence of an island of extreme
nuclear isomerism arising from core-excited configurations 
at high excitation in the vicinity of the doubly closed-shell 
nucleus $^{208}$Pb. These results are expected to provide 
discriminating tests of the effective interactions used 
in current large-scale shell-model calculations. 
\end{abstract}


\maketitle

Metastable states in atomic nuclei, also referred to as isomers, represent 
the manifestation of the associated wave functions being pure and quite distinct from 
those of other levels in their vicinity. Consequently, transition
rates for decay of these states are orders of magnitude lower than those
of nearby levels. The exploration of a variety of nuclear isomers, 
whose decay may be hindered by the required large change in angular 
momentum, or in its projection on the symmetry axis of a deformed nucleus, 
or in its shape, or a considerable difference in the configurations of initial
and final levels, leads to crucial insights which further the understanding of 
the strongly-interacting, nuclear many-body system. Specifically, isomeric
properties play a major role in refining effective interactions 
for shell-model calculations of near-spherical nuclei. Detailed descriptions
are available in recent reviews \cite{Dracoulis2016,Walker1999,Kondev2015,Walker2020}. 
In some instances, the degree of hindrance of the decay may be quite 
extreme, leading to isomeric half-lives which are larger by many orders 
of magnitude in comparison to those of other similar states. Some examples of
such extreme nuclear isomerism are the: (a) ``spin isomer" 
in $^{180}$Ta ($Z$=73) with {\it T}$_{1/2}$ $>$ 7.1 x 10$^{15}$ y \cite{Hult2006}, 
(b) ``{\it K} isomer" in $^{178}$Hf ($Z$=72) with {\it T}$_{1/2}$ = 31(1) y \cite{Helmer1973}, 
(c) ``shape isomer" in $^{242}$Am ($Z$=95) with {\it T}$_{1/2}$ = 14(1) ms 
\cite{Polikanov1962}. All of these isomers lie at relatively low ($<$ 2.5 MeV) 
excitation energy. In the region around the doubly closed-shell nucleus
$^{208}$Pb, a notable isomer at relatively low excitation is
the $\alpha $-decaying 2.93-MeV state with {\it T}$_{1/2}$ = 45.1(6) s 
in $^{212}$Po \cite{Perlman1962}. With an increase in excitation energy, a trend of 
decreasing half-lives is evident. In lighter nuclei, with fewer valence nucleons
and lower level density, in some instances, longer-lived states
may result, such as the 45.9(6)-s level at 6.958 MeV, with total angular 
momentum (henceforth, referred to as spin), {\it I} = 12 $\hbar $, in $^{52}$Fe 
($Z$=26) \cite{Geesaman1975}, and the $\beta $-decaying
(21$^{+}$), 0.40(4)-s state at 6.67 MeV in $^{94}$Ag \cite{LaCommara}. 
The longest-lived isomers at very high excitation ($>$ 7 MeV), 
known prior to this work, are the 8.533-MeV state in $^{212}$Fr ($Z$=87) 
with {\it T}$_{1/2}$ = 23.6(21) $\mu $s \cite{Byrne1990}, 
and the 8.095-MeV level in $^{213}$Fr with {\it T}$_{1/2}$ = 3.1(2) 
$\mu $s \cite{Byrne1989}. In fact, the isomer in $^{212}$Fr had been
characterized as an ``outstanding example" of a spin trap in
near-spherical nuclei, and given its long half-life and high excitation
energy, had been termed as an ``extreme isomer" \cite{Walker1999}. It had
also been recognized then that ``more extreme isomers might exist in heavy 
nuclei" \cite{Walker1999}. It should be noted though that no specific
predictions were made. The isomers reported in the present work have
been discovered two decades later, despite the large body of recent
work on isomers in the {\it A} = 200-215 region by several groups worldwide 
\cite{Kondev2021}. The presence of such long-lived states in the $>$7-8 MeV 
excitation range, which decay by $\gamma $-ray emission, is noteworthy.

Surveys of isomers across the nuclear chart \cite{Kondev2021,Jain2015} 
reveal that many long-lived states at high excitation are known to exist 
in heavy nuclei, which lie near the line of $\beta $-stability. As a result,
these isomers are difficult to access experimentally, since compound
nuclear fusion-evaporation reactions, which can populate levels at the
highest excitation, favor the production of isotopes deficient in 
neutrons. Inelastic excitation and multi-nucleon transfer reactions,
on the other hand, can be used to access nuclei near the line of
stability or even on the neutron-rich side. However, the cross sections and highest spin 
attainable are limited in comparison with fusion-evaporation products.
With the sensitivity provided by large $\gamma $-ray detector arrays,
and pulsed beams from accelerators with a range of timing options,
exploring isomers at high excitation near the line of $\beta $-stability
becomes feasible. An impressive example is the recent study of 
the heaviest known doubly closed-shell nucleus $^{208}$Pb ($Z$=82, $N$=126),
wherein states up to spin 30 $\hbar $ and excitation energy, {\it E}$_{x}$ = 16.4 MeV,
have been established \cite{Broda2017}.
The focus of the present work is the region near $^{208}$Pb,
where the availability of numerous orbitals with high values of
intrinsic angular momentum, {\it e.g.}, {\it h}$_{11/2}$
proton ($\pi $) and {\it i}$_{13/2}$ neutron ($\nu $) holes, and $\pi ${\it h}$_{9/2}$, 
$\pi ${\it i}$_{13/2}$ and $\nu ${\it g}$_{9/2}$ particles,
leads to conditions conducive for the realization of long-lived
states at very high excitation. Recent work, including that of this collaboration
\cite{Wahid2020,Roy2019,Bothe2022}, has revealed the presence 
of several states with half-lives
up to hundreds of microseconds at intermediate excitation, but the 
23.6(21)-$\mu $s isomer in $^{212}$Fr at 8.533 MeV was thus far the 
longest-lived state above 7 MeV \cite{Byrne1990}. The present work describes newly
identified metastable states in $^{205}$Bi ({\it Z} = 83) and $^{204}$Pb whose
half-lives are about two orders of magnitude larger than other 
isomers in a similar excitation range. Additionally, the half-life
of the 10.17-MeV state in $^{206}$Bi, which was previously reported
to be $>$ 2 $\mu $s \cite{Cieplicka2012}, has been measured and found to be slightly
longer than that of the $^{212}$Fr isomer. These newly-identified metastable states 
have a different character as compared to the so-called ``spin isomers" 
in the Rn-Fr-Ra region, as will be described below. 

The work described in this letter involves the population of 
highly-excited levels in isotopes of Pb, Bi and other elements
through multi-nucleon transfer reactions 
with heavy, highly-energetic projectiles:
(a) 1450-MeV $^{209}$Bi and (b) 1430-MeV $^{207}$Pb beams, incident on
a 50 mg/cm$^{2}$ Au target. The Gammasphere detector array \cite{LeeJanssens}, 
which at the time consisted of 100 Compton-suppressed high-purity
germanium detectors, was used to record coincident $\gamma $ rays
emitted within $\approx $1 $\mu $s of each other. Details regarding 
the experiment and data analysis are presented elsewhere \cite{Tandel2015}.
In previous reports on $^{205}$Bi and $^{204}$Pb, levels up to 6.7 MeV
and 8.1 MeV, respectively, had been identified utilizing $\alpha $
particles as projectiles incident on $^{205}$Tl and $^{204}$Hg targets
\cite{Byrne1989_2,Linden1978}. These experiments were focused on the 
decays of short-lived levels, and a number of transitions up to
{\it I} $\approx $ 20 $\hbar $ were placed in the respective level
schemes. The focus of the present work was on identifying metastable
states and establishing their half-lives and decay paths.

Pulsed beams from the ATLAS accelerator at Argonne National Laboratory
were used in different beam-sweeping intervals: successive bursts
separated by $\approx $ 825 ns up to 8 s, enabled a search for and
identification of isomers with half-lives in the microseconds,
milliseconds and seconds time ranges. The data were collected in
``beam-off" periods of 800 $\mu $s, 3 ms, 3 s and 8 s, {\it i.e.},
the pulsed beam was deflected away from the target for these durations.
The coincidence window was $\approx $ 1 $\mu $s. When beam pulses
were separated by 825 ns, three- or higher-fold coincidence data were
collected. For larger time periods, during the ``beam-off" periods of
800 $\mu $s and above, two- and higher-fold data were recorded.
The data were sorted offline into histograms of two, three and four
dimensions involving energy and time parameters, and subsequently
analyzed using the RADWARE and TSCAN suite of programs
\cite{Radford1995,Jin1992}. Some examples of the histograms used
for the data analysis are listed here:
(a) two-, three- and four-dimensional symmetric, $\gamma $-energy
histograms for establishing the excited level structures;
(b) time-gated, triple-$\gamma $ energy coincidence histograms to
establish long half-lives;
(c) energy-energy-time difference histograms for determining
half-lives {\it T}$_{1/2}$ $<$ 1 $\mu $s;
(d) prompt-delayed, two- and three-dimensional $\gamma $-energy
histograms for identifying coincidence events across isomeric
states with {\it T}$_{1/2}$ $<$ 1 $\mu $s;
(e) angle-sorted, $\gamma $-energy, asymmetric matrices to
determine transition multipolarities using the method of directional
angular correlations from oriented states (DCO) \cite{Krane73}.
The so-called ``prompt" and ``delayed" coincidence events
corresponded to the detection of at least three $\gamma $ rays
within $\pm $40 ns and 50-650 ns of the trigger, respectively,
when the beam pulses were separated by 825 ns. 

A total of thirty new $\gamma $ rays have been placed
in the level schemes of $^{204}$Pb and $^{205}$Bi from the present work. 
However, only the transitions crucial for establishing
the spin and parity quantum numbers of the isomeric levels are
discussed. A paper describing the detailed level schemes for
$^{204}$Pb and $^{205}$Bi deduced from this work is being prepared 
\cite{Wahid2022}. The newly established $\gamma $ rays, along with
previously reported ones, are displayed in Fig. 1.
The $\gamma $-ray spectra illustrated in Figs. 1 and 2
represent three-fold delayed coincidence events. In Figs. 1(a) and 1(b),
the summed coincidence spectra obtained with two simultaneous gates
on all combinations of pairs of transitions in the previously known
cascades in: (a) $^{204}$Pb: 1006-325-618-167-1046-316-433 keV spanning
{\it I} = 9 to 19 $\hbar $, and (b) $^{205}$Bi: 881-286-697-641-600-516 keV
spanning {\it I} = 9/2 to 31/2 $\hbar $, are presented. The quality of 
the data with fewer gating transitions, specifically the three decay 
branches from the isomer in $^{204}$Pb, are displayed in Fig. 2.

The level scheme of $^{204}$Pb has been extended up to a new
long-lived isomer at {\it E}$_{x}$ = 8349 keV, the half-life of which 
has been determined to be 0.22(2) ms [Fig. 3(a)], by inspecting the 
time distribution of the summed coincidence counts in the previously known 
1006-325-618-167-1046-316-433 keV cascade spanning {\it I} = 9-19 
$\hbar $ \cite{Linden1978}, when the 
beam was incident on the target for 200 $\mu $s, and data were
collected during a 800-$\mu $s beam-off period. In the case of
$^{205}$Bi, an inspection of the 800-$\mu $s and 3-ms beam-off data 
indicated that the half-life of the isomer was greater than 
these periods. Therefore, data from 
the next higher available pulsing period (3-s beam-off) were
scrutinized. The time distribution of the summed coincidence
counts in the previously known 641-600-516-keV cascade [Fig. 1(b)] 
spanning a spin range 25/2 to 31/2 $\hbar $ \cite{Byrne1989} indicate a
half-life of 8(2) ms for the metastable state in $^{205}$Bi 
[Fig. 3(b)]. To determine the half-life of the 10.17-MeV level
in $^{206}$Bi, previously reported to be $>$ 2 $\mu $s \cite{Cieplicka2012}, 
the time distribution of the summed coincidence counts of the most intense 
$\gamma $ rays in the cascade between the 10.17 and 1.045 MeV 
levels was inspected in the 800-$\mu $s beam-off data,
leading to a half-life of 0.027(2) ms, as 
indicated in Fig. 3(c). A comparison of the above half-lives with
those of other isomeric levels above an excitation energy of 7 MeV
\cite{Jain2015}, and with {\it T}$_{1/2}$ $>$ 1 $\mu $s,
across the nuclear chart is displayed in Fig. 4, where it is
evident that the data points for $^{204}$Pb and $^{205}$Bi are
outliers compared to those previously identified. 

The excitation energy and spin-parity quantum numbers for the 
newly-identified isomer in $^{204}$Pb were established as 
{\it E}$_{x}$ = 8349 keV and {\it I}$^{\pi }$ = (22$^{+}$).
The 481-keV $\gamma $ ray in $^{204}$Pb (Fig. 1) is not
observed in the so-called ``prompt" data, which are recorded within
a few tens of nanoseconds of the beam being incident on the target,
but is clearly visible in the data collected during the ``beam-off"
periods. Therefore, it is attributed to the direct deexcitation of the
8349-keV isomer. The 481- and 2520-keV $\gamma $ rays, which are newly
identified in the present work, are found to be in cascade,
with the latter directly feeding the previously established 16$^{+}$, 
5348-keV level with a proposed $\nu(i^{-2}_{13/2}, f^{-1}_{5/2}, p^{-1}_{3/2})$ 
configuration in $^{204}$Pb \cite{Linden1978}. The 2520-keV transition most likely
has $E3$ character based on a calculation following the prescription in
previous work \cite{Wrzesinski2003,Broda2011,Szpak2011,Wrzesinski2015}, described below,
leading to an {\it I}$^{\pi }$ = (19$^{-}$) assignment for the $E_x$ = 7868-keV
initial level. The expected transition energy for the $E3$ excitation built
on the four-nucleon-hole, 16$^{+}$ state can be estimated as follows.
The unperturbed energy of the $E3$ excitation in $^{208}$Pb would be
2615 keV. On account of the coupling to configurations involving multiple
nucleons, energy shifts would result from the two {\it i}$_{13/2}$ neutrons
and the two low-{\it j} neutrons.
The final energy can be expressed as the sum of energy shifts
corresponding to the individual constituents of such a configuration,
which, in this case, turns out to be 2483 keV, in fair agreement with the
experimentally observed 2520-keV value, thus validating its $E3$ assignment.
To determine the multipolarity of the 481-keV transition
feeding the level deexcited by the 2520-keV $\gamma $ ray,
intensity balance considerations have been used,
for which a detailed procedure may be found in our earlier work
\cite{Wahid2020,Roy2019,Bothe2022}: either $E$3 or $M$1 character
is inferred, due to the similarity of the theoretical total
conversion coefficients (0.111 and 0.119, respectively), from
BRICC \cite{Kibedi2008}. Based on typical transition rates expected for
$\gamma $ rays with different multipolarities, $M$1 character appears
unlikely.  An $E$3 character for the 481-keV transition would imply
spin-parity quantum numbers (22$^{+}$) for the isomeric level.
This would be consistent with similar isomeric transitions in
$^{212,213}$Fr, $^{206}$Bi \cite{Byrne1990,Byrne1989,Cieplicka2012},
and many other nuclei in the vicinity of $^{208}$Pb. It may be noted
that there are no long-lived isomers in this region which
decay via $M$1 transitions.

In $^{205}$Bi, it was only possible to constrain the spin-parity 
of the isomer based on various experimental and theoretical 
considerations, but a firm assignment, comparable to the $^{204}$Pb 
case, was not possible. All $\gamma $ rays assigned to $^{205}$Bi [Fig. 1(b)]
are visible in the ``prompt" data ruling out the possibility 
that any of these directly deexcite the isomer. The 8-ms isomer, therefore, 
most likely deexcites through one or more unobserved low-energy transitions 
with large conversion coefficients, accounting for their absence in the 
spectra. Similar considerations, as those outlined above for $^{204}$Pb, 
have been employed in the case of $^{205}$Bi,
where the 2442-482-295 keV cascade feeds the previously established
4696-keV, 37/2$^{-}$ level \cite{Byrne1989_2}. From the
present work, the energy of this level is inferred to be about 1 keV
lower {\it i.e.}, 4695 keV. The 2442-keV $\gamma $ ray feeds this level,
and a similar calculation as in $^{204}$Pb leads to the inference of its
$E3$ character. Therefore, the level at 7136 keV deexcited by the
2442-keV $\gamma $ ray is assigned {\it I}$^{\pi }$ = (43/2$^{+}$).
Intensity balance considerations imply $M1$ character for the 295-keV
transition. The 482-keV $\gamma $ ray may have either $M1$ or $E3$
character, with the latter being excluded by the prompt decay of the
associated level. The 2442-482-295 keV cascade is therefore assigned
E3-M1-M1 multipolarity, with the topmost level at 7913 keV having
possible {\it I}$^{\pi }$ = (45/2$^{+}$) or (47/2$^{+}$). The 8-ms
isomer in $^{205}$Bi probably decays to this level through an unobserved
transition. It may be noted that there is another level at 7971 keV
in $^{205}$Bi which has a prompt decay, and is fed by the decay of
the isomer through an unobserved transition. It is not clear from
the data whether the isomer in $^{205}$Bi has two decay paths: one
each to the levels at 7913 and 7971 keV, or only one decay to
the 7971-keV state, which then deexcites to the 7913-keV level.
All of the above considerations will be described in our detailed
future paper \cite{Wahid2022}. Regarding the excitation
energy of the 8-ms isomer in $^{205}$Bi, a value 7913+$x$ keV is 
listed, where $x$ $<$ 150 keV. This upper limit has been
estimated taking into account the statistics in the data, inferred
transition probabilities, theoretical conversion coefficients of
the expected $E3$ or $M2$ transitions \cite{Kibedi2008}, and the
efficiency of the detector array. A transition of 150 keV would
have theoretical total conversion coefficients of 18.9 and 18.8
for $M2$ and $E3$ multipolarities, respectively. The available
statistics in the data and the efficiency of the array at 150 keV
would imply 2-3 counts for such a $\gamma $ ray, with a background
of about 1 count. The resultant peak may or may not be discernible
in the spectra, therefore the upper limit of 150 keV. Of course,
it is quite possible that the actual energy or energies of the
transitions deexciting the isomer is significantly lower,
which is why $x$ $<$ 150 keV represents an upper limit only. We have
performed additional calculations towards a more sophisticated
estimate which will be described in our future paper
\cite{Wahid2022}. In view of the measured half-life and the 
corresponding inferred transition rates for different multipolarities, 
and based on the results of shell-model calculations performed (described below), 
a (51/2$^{-}$) spin-parity assignment appears quite probable for the isomer
implying a decay through low-energy ($<$ 150 keV) $E$3 and/or 
$M$2 transitions. 

Prior to this work, metastable states reported in this region beyond 
{\it E}$_{x}$ = 7 MeV, with the highest half-lives, were in the 
Rn-Fr-Ra isotopes \cite{Dracoulis2016,Walker1999,Byrne1990,Byrne1989}.
The ones identified in Tl-Pb-Bi isotopes prior to this work were primarily at lower 
excitation \cite{Wrzesinski2003,Broda2011,Szpak2011,Wrzesinski2015,Broda2018},
except the states at very high excitation in $^{208}$Pb \cite{Broda2017}.
The primary differences in the nature of the isomers in the above two
regions are evident from an inspection of: (i) excitation energy ($E_x$)
as a function of $I(I+1)$, where $I$ is the spin, and 
(ii) reduced $E$3 transition probabilities [$B(E3)$] for the decay of these isomers. 
As demonstrated in Fig. 5(a), the {\it I}$^{\pi }$ = (22$^{+}$) isomeric 
state in $^{204}$Pb follows the trajectory of the lowest-energy levels at 
lower spin, as is also the case for others in neighboring nuclei. However, the
{\it I}$^{\pi }$ = (34$^{+}$) isomeric level in $^{212}$Fr [Fig. 5(b)]
is found to lie distinctly lower in energy at the given spin in relation
to the trajectory defined by the lower-lying states, attesting to its
nature as a ``spin-trap" isomer, similar to the situation in 
neighboring isotopes of Rn, Fr and Ra. 
Another striking difference is evident in the $B(E3)$ values for the decay 
of the isomers in these two regions [Figs. 5(c) and 5(d)]. In the Tl-Pb-Bi
region, including the isomers newly identified from this work, the 
$B(E3)$ values are found to be in the vicinity of those expected from
Weisskopf single-particle transition rates [Weisskopf units (W.u.)], 
or to be significantly lower. In the Rn-Fr-Ra region, an
increase ranging from 20-40 units in comparison to the single-particle
transition rates is visible in the $B(E3)$ values. This is attributed 
to transitions involving a change $\Delta ${\it j} = $\Delta ${\it l} = 3 
between states \cite{Dracoulis2016}. The relatively
lower half-lives in this region can be explained in terms of the
resultant enhancement in the $E$3 transition probabilities. In the
Tl-Pb-Bi region though, no such enhancement is visible, and for
the isomers reported in the present work, the half-lives are 
considerably larger. It is relevant to mention here that $E3$ decays 
from the 23$^{+}$ and 28$^{-}$ isomers at 11.4 MeV and 13.7 MeV in 
$^{208}$Pb have been observed to be enhanced by factors of 32 and 56 
W.u., respectively \cite{Broda2017}, similar to the 15/2$^{-}$ 
$\rightarrow $ 9/2$^{+}$ transition in $^{209}$Pb which has a strength 
of 26(7) W.u. This enhancement is similar to that observed 
in the Rn-Fr-Ra region and may be attributed to the very high excitation 
in $^{208}$Pb which makes it possible to sample the {\it j}$_{15/2}$ 
neutron orbital, resulting in transitions of the  $\nu ${\it j}$_{15/2}$
$\rightarrow $ $\nu ${\it g}$_{9/2}$ type. At low spin, in $^{209}$Pb,
the 15/2$^{-}$ level is 1.42 MeV above the 9/2$^{+}$ ground state, with these 
levels resulting from the occupation of the {\it j}$_{15/2}$ and {\it g}$_{9/2}$
orbitals, respectively, therefore the half-life of the 15/2$^{-}$ level
is only 1.36(30) ns \cite{NDS209}. 
It is possible that long-lived states at high excitation may also be 
realized in Po ({\it Z} = 84) and At ({\it Z} = 85) isotopes
with $N < 126$, and neutron-rich ones with {\it Z} $<$ 82,
which would become accessible with rare-isotope beams. 

It may be noted that the isomers in the
Rn-Fr-Ra region were populated through fusion-evaporation reactions,
as compared to multi-nucleon transfer reactions in the present work. 
In the latter case, quality spectroscopic data at high excitation
are relatively more difficult to obtain, therefore long-lived  
isomers in this energy range in the Tl-Pb-Bi region 
around the line of stability remained undiscovered until now. It is
noteworthy that the isomers in the Tl-Pb-Bi region involve either 
hole-hole or particle-hole excitations, and in the Rn-Fr-Ra case particle-particle
configurations, comprising nucleons in high-{\it j} orbitals.  

To aid in the understanding of the experimental data, shell-model calculations 
have been performed for $^{204}$Pb and $^{205}$Bi using the KHH7B 
effective interaction in the model space $Z$ = 58-114 and $N$ = 100-164 
around doubly-magic $^{208}$Pb using the OXBASH code \cite{Brown2004}.
The model space includes the proton orbitals $1d_{5/2}$, $0h_{11/2}$, 
$1d_{3/2}$ and $2s_{1/2}$ below $Z=82$, and the $0h_{9/2}$, $1f_{7/2}$,
and $0i_{13/2}$ ones above, and the neutron orbitals $0i_{13/2}$, $2p_{3/2}$,
$1f_{5/2}$, and $2p_{1/2}$ below $N=126$ and the $1g_{9/2}$,
$0i_{11/2}$, and $0j_{15/2}$ ones above it. For the KHH7B effective interaction,
the cross-shell two-body matrix elements (TBMEs) are taken from the
G-matrix potential (H7B) \cite{Hosaka1985}, while the proton-neutron
hole-hole and particle-particle TBMEs are from the Kuo-Herling interaction
\cite{KuoH1971}, with modifications included later \cite{WarBrown1991}.
For $^{204}$Pb, two sets of calculations were performed with $t=0$ and
$t=1$, where $t=0$ represents no excitation across the $Z=82$ and $N=126$
shell gaps. In the $t=1$ case, the calculations involve the excitation
of one nucleon across the $Z=82$ and $N=126$ shell gaps. Mixing between
$t=0$ and core-excited configurations is blocked in these calculations.
For $^{205}$Bi, the $1d_{5/2}$, $0h_{11/2}$, $1d_{3/2}$ and $2s_{1/2}$
proton orbitals are completely filled, with the unpaired proton occupying
either the $0h_{9/2}$, $1f_{7/2}$ or $0i_{13/2}$ orbitals. The 22 valence
neutrons have been allowed to occupy only the $0i_{13/2}$, $2p_{3/2}$,
$1f_{5/2}$ and $2p_{1/2}$ orbitals below $N=126$. Results of shell-model
calculations using the KHH7B interaction for other nuclei in this region
have been recently reported \cite{Wahid2020,Wilson2015,Berry2019,Anil2021}.

In $^{204}$Pb, the shell-model calculations indicate the presence
of multiple $t=1$ states with spin-parity quantum numbers 19$^{-}$
and 20$^{-}$, and excitation energy $\approx $ 7-8 MeV. A
20$^{+}$ state with a $\nu i^{-4}_{13/2}$ configuration is also
expected to lie in this vicinity. It is quite unlikely that any
of these states are candidates for the isomer at 8349 keV since,
in that case, the 481-keV deexciting transition (Fig. 1) would
be of dipole character and would be inconsistent with the 0.22(2) ms half-life.
The calculations indicate that states with spin $>$ 20 $\hbar $
can arise in either of two ways {\it viz}., the
$\pi(h^{-1}_{11/2}$, $h_{9/2})\otimes \nu (i^{-2}_{13/2})$ and the 
$\pi(h^{-1}_{11/2}$, $h_{9/2})\otimes \nu (f^{-1}_{5/2}$, $p^{-1}_{1/2}$, $i^{-2}_{13/2})$ 
configurations. While both these configurations can lead to the 
22$^{+}$ level calculated to be at 8085 keV, the amplitude for the 
4-quasiparticle state is found to be 14.2$\%$, while for the one with 
six quasiparticles it is 56.1$\%$. Though the excitation energies
are reasonably reproduced in most cases, the shell-model calculations
do not give a good account of the measured transition probabilities.
 
In $^{205}$Bi, levels with spin-parity quantum numbers 43/2$^{-}$,
45/2$^{+}$, 45/2$^{-}$, 47/2$^{+}$, 47/2$^{-}$ and 51/2$^{-}$ are 
calculated to lie in the region between 6.5-7.7 MeV. It is quite 
unlikely that the isomer at an excitation energy of 7913+$x$ keV 
has {\it I} $\le $ 47/2 $\hbar $ since, in that case, relatively 
fast $E$2 or $M$2 transitions of several hundred keV would 
deexcite to levels with spin up to 43/2 $\hbar $, 
inconsistent with the measured 8-ms half-life. While 
the shell-model calculations indicate that multiple 51/2$^{-}$ 
states are possible, only one of these, with the
$\pi i_{13/2} \otimes \nu(f^{-1}_{5/2}, i^{-3}_{13/2})$
configuration, is low enough in energy to be consistent with the
experimental value. Four other 51/2$^{-}$ states, all
involving the $\nu g_{9/2}$ orbital, are possible but 
are calculated to lie above 9 MeV, and are therefore unlikely to 
be candidates for the isomeric configuration. Further, all other levels such as 
the 53/2$^{+}$ state with the $\pi i_{13/2} \otimes \nu(i^{-4}_{13/2})$ 
configuration are also unlikely since they lie above 9 MeV.
Therefore, based on the expectation from the shell-model calculations,
the only candidate for the isomer would be the 51/2$^{-}$ state with the
$\pi i_{13/2} \otimes \nu(f^{-1}_{5/2}, i^{-3}_{13/2})$ configuration.

In the previous work on $^{206}$Bi \cite{Cieplicka2012}, configuration 
assignments for the isomers with {\it I}$^{\pi }$ = (28$^{-}$) and 
(31$^{+}$), and half-lives of 155 ns and 0.027 ms, respectively, had
not been proposed. The (28$^{-}$) isomer likely results from the 
$\pi i_{13/2} \otimes \nu  [i_{13/2}^{-3}, (p_{1/2}^{-1},g_{9/2})]_{43/2-}$ 
configuration, with the (26$^{+}$) state to which it decays having a 
$\pi h_{9/2} \otimes \nu  [i_{13/2}^{-3}, (p_{1/2}^{-1},g_{9/2})]_{43/2-}$ 
one. Thus, the isomerism would be associated with the $\pi i_{13/2}$ 
$\rightarrow $ $\pi h_{9/2}$ transition with a strength $B(M2)$ = 0.028(3) W.u.,
consistent with hindered $M$2 decays in this region. 
There are two possibilities for the configuration of the (31$^{+}$) isomer: 
either a neutron core excitation, $\pi i_{13/2} \otimes \nu  [i_{13/2}^{-3}, (p_{1/2}^{-1},j_{15/2})]_{49/2+}$,
or a proton core excitation, $\pi [i_{13/2},(h_{11/2}^{-1},h_{9/2})] \otimes \nu  (i_{13/2}^{-3})$.
The measured half-life leads to $B(E3)$ = 0.0276(21) W.u., consequently 
the latter scenario (proton core excitation) seems more likely, since
in the former instance, the $\nu j_{15/2} \rightarrow \nu g_{9/2}$,
$\Delta ${\it j} = $\Delta ${\it l} = 3, $E$3 transition would be expected to 
have a transition probability $B(E3)$ $>$ 20 W.u., far from the value deduced 
from the present data. 

The new discoveries from the present work have direct implications for
understanding the structure of nuclei in the vicinity of the heaviest, 
doubly-magic nucleus, $^{208}$Pb. The identification of isomers at high 
excitation with complex core-excited configurations, where the excitation
energy and spin-parity are established firmly from experiment,
challenges large-scale shell-model calculations using available 
interactions. Additionally, with a precise knowledge of the half-lives of
the isomers and the decay paths, tests of the predicted 
electromagnetic properties are also feasible. Further, long-lived 
isomeric states in general, and particularly in the present cases, tend to 
have relatively pure configurations, and are, therefore, more suitable to
discriminate between different theoretical predictions. As stated earlier,
the character of the isomers at high spin in the Tl-Pb-Bi region is different 
from that of the long-lived states previously established in the Rn-Fr-Ra one 
(which are yrast spin-traps, and have enhanced $E3$ transition rates). Therefore, 
different considerations have to be factored in their respective descriptions. 
At present, the commonly used effective interactions in this region are the: 
(a) KHHE ({\it Z} = 58-82, {\it N} = 82-126),  
(b) KHH7B ({\it Z} = 58-114, {\it N} = 100-164),  
(c) KHPE ({\it Z} = 82-126, {\it N} = 126-184),  
(d) $V_{low-k}$ (Z = 82-126, N = 126-184), and 
(e) KHM3Y ({\it Z} = 50-126, {\it N} = 82-184).  
It is recognized that, while these interactions have satisfactory predictive
power at low to intermediate excitation, there are major limitations in the 
description of high-spin phenomena, as for example, outlined in the
previous work on $^{206}$Bi \cite{Cieplicka2012}. The present results, 
{\it viz.}, the excitation energy, spin-parity and electromagnetic
properties of the three longest-lived metastable states at high excitation 
observed thus far in the periodic chart constitute an important contribution 
which should improve our understanding considerably. 

To summarize, metastable states have been established in 
$^{205}$Bi, $^{204}$Pb and $^{206}$Bi, with half-lives of 8(2) ms, 0.22(2) ms 
and 0.027(2) ms, respectively, constituting the three highest values of 
half-lives above an excitation energy of 7 MeV across the nuclear chart.   
This suggests that nuclear isomerism built on 
core-excited configurations is quite robust even under 
such extreme conditions. The emergence of a new frontier of metastable states 
in nuclei with {\it T}$_{1/2}$ $\gg$ 1 $\mu $s at high excitation, with the
hindrance attributable to the difference in the configurations of isomeric
states and the levels to which they decay, close to the 
line of $\beta $-stability in the region around $^{208}$Pb, is 
indicated. These results constitute a fertile testing ground for large-scale
shell-model calculations, along with available effective interactions. 
With focused experiments to identify more such instances, using multi-nucleon 
transfer reactions with highly-energetic, heavy-ion beams and suitably 
long pulsing periods, coupled with the sensitive detection of $\gamma $ rays 
using large detector arrays, which are being planned by this collaboration,
similar long-lived states are expected to be found in neighboring nuclei,
thus redefining one extreme of isomerism.  

The authors would like to thank I. Ahmad, J.P. Greene, A.J. Knox, D. Peterson, 
X. Wang and C.M. Wilson for assistance during the experiment, 
and M. Hemalatha for insightful comments. 
S.G.W. acknowledges support from the DST-INSPIRE Ph.D. Fellowship of the Department 
of Science and Technology, Government of India (Fellowship No. IF150098); 
S.K.T. from the University Grants Commission, India, under the Faculty Recharge 
Programme, and S.S. from the DST-INSPIRE Ph.D. Fellowship of the Department 
of Science and Technology, Government of India (Fellowship No. IF170965). 
P.C.S. acknowledges a research grant from SERB (India), CRG/2019/000556.
This work is supported by the U.S. Department of Energy, Office of Science, 
Office of Nuclear Physics, under award numbers DE-FG02-94ER40848, 
DE-FG02-94ER40834 (UML), DE-FG02-97ER41041 (UNC) and DE-FG02-97ER41033 (TUNL), 
and contract number DE-AC02-06CH11357 (ANL). The research described 
here utilized resources of the ATLAS facility at ANL, which is a DOE 
Office of Science user facility.

\bibliographystyle{amsplain}

\begin{thebibliography}{35}

\bibitem{Dracoulis2016} G.D. Dracoulis {\it et al}., Rep. Prog. Phys. {\bf 79}, 076301 (2016),
and references therein.

\bibitem{Walker1999} P.M. Walker and G.D. Dracoulis, Nature (London) {\bf 399}, 35 (1999).

\bibitem{Kondev2015} F.G. Kondev {\it et al}., Atomic Data and Nuclear Data Tables {\bf 103-104}, 50-105 (2015).

\bibitem{Walker2020} Philip Walker and Zsolt Podolyak, Phys. Scr. {\bf 95}, 044004 (2020).

\bibitem{Hult2006} Mikael Hult {\it et al}., Phys. Rev. C. {\bf 74}, 054311 (2006).

\bibitem{Helmer1973} R.G. Helmer and C.W. Reich, Nucl. Phys. A {\bf 211}, 1 (1973).

\bibitem{Polikanov1962} S.M. Polikanov {\it et al}., Soviet Phys. JETP {\bf 15}, 1016 (1962).

\bibitem{Perlman1962} I. Perlman {\it et al}., Phys. Rev. {\bf 127}, 917 (1962).

\bibitem{Geesaman1975} D.F. Geesaman {\it et al}., Phys. Rev. Lett. {\bf 34}, 326 (1975).

\bibitem{LaCommara} M. La Commara {\it et al}., Nucl. Phys. A {\bf 708}, 167 (2002).

\bibitem{Byrne1990} 
A.P. Byrne {\it et al}., Phys. Rev. C {\bf 42}, R6 (1990). 

\bibitem{Byrne1989}
A.P. Byrne {\it et al}., Phys. Lett. B {\bf 217}, 38 (1989).

\bibitem{Kondev2021}
F.G. Kondev {\it et al}., Chinese Phys. C 45 030001 (2021).

\bibitem{Jain2015}
A. K. Jain {\it et al}., Nucl. Data Sheets {\bf 128}, 1 (2015). 

\bibitem{Broda2017}
R. Broda {\it et al}., Phys. Rev. C {\bf 95}, 064308 (2017).

\bibitem{Wahid2020}
S.G. Wahid {\it et al}., Phys. Rev. C {\bf 102}, 024329 (2020). 

\bibitem{Roy2019}
Poulomi Roy {\it et al}., Phys. Rev. C {\bf 100}, 024320 (2019).

\bibitem{Bothe2022}
V. Bothe {\it et al}., Phys. Rev. C (2022).

\bibitem{Cieplicka2012}
N. Cieplicka {\it et al}., Phys. Rev. C {\bf 86}, 054322 (2012).

\bibitem{LeeJanssens}
I-Yang Lee, Nucl. Phys. A {\bf 520}, c641 (1990), and R.V.F. Janssens
and F.S. Stephens, Nucl. Phys. News {\bf 6}, 9 (1996).

\bibitem{Tandel2015}
S.K. Tandel {\it et al}., Phys. Lett. B {\bf 750}, 225 (2015). 

\bibitem{Byrne1989_2}
A.P. Byrne {\it et al}., Z. Physik A {\bf 334}, 247 (1989).

\bibitem{Linden1978}
C.G. Linden {\it et al}., Z. Physik A {\bf 284}, 217 (1978).

\bibitem{Radford1995}
D.C. Radford, Nucl. Inst. Meth. A {\bf 361}, 297 (1995).

\bibitem{Jin1992}
H.-Q. Jin, TSCAN and related programs, RUTGERS-ORNL-UTK, 1992-1997.

\bibitem{Krane73}
K.S. Krane {\it et al}., Nucl. Data Tables 11 (1973) 351.

\bibitem{Wahid2022}
S.G. Wahid {\it et al}., to be published.

\bibitem{Wrzesinski2003}
J. Wrzesinski, Eur. Phys. J. A {\bf 20}, 57 (2003).

\bibitem{Broda2011}
R. Broda {\it et al}., Phys. Rev. C {\bf 84}, 014330 (2011).

\bibitem{Szpak2011}
B. Szpak {\it et al}., Phys. Rev. C {\bf 83}, 064315 (2011).

\bibitem{Wrzesinski2015}
J. Wrzesinski {\it et al}., Phys. Rev. C. {\bf 92}, 044327 (2015). 

\bibitem{Kibedi2008}
T. Kibedi {\it et al}., Nucl. Instr. Meth. A {\bf 589}, 202 (2008).

\bibitem{Broda2018}
R. Broda {\it et al}., Phys. Rev. C. {\bf 98}, 024324 (2018). 

\bibitem{NDS209}
J. Chen, F.G. Kondev, Nucl. Data Sheets {\bf 126}, 373 (2015).

\bibitem{Brown2004}
 B.A. Brown {\it et al}., OXBASH for Windows, MSU-NSCL report 1289, 2004.

\bibitem{Hosaka1985}
A. Hosaka {\it et al}., Nucl. Phys. A {\bf 444}, 76 (1985).

\bibitem{KuoH1971}
T.T.S. Kuo and G.H. Herling, Report No. 2258, US Naval Research Laboratory,
1971, unpublished.

\bibitem{WarBrown1991}
E.K. Warburton {\it et al}., Phys. Rev. C {\bf 43}, 602 (1991).

\bibitem{Wilson2015}
E. Wilson {\it et al}., Phys. Lett. B {\bf 747}, 88 (2015).

\bibitem{Berry2019}
T.A. Berry {\it et al}., Phys. Lett. B {\bf 793}, 271 (2019).

\bibitem{Anil2021}
A. Kumar and P.C. Srivastava, Nucl. Phys. A {\bf 1014}, 122255 (2021).

\end{thebibliography}

\clearpage

\begin{figure}
\begin{center}
\includegraphics[scale=0.4]{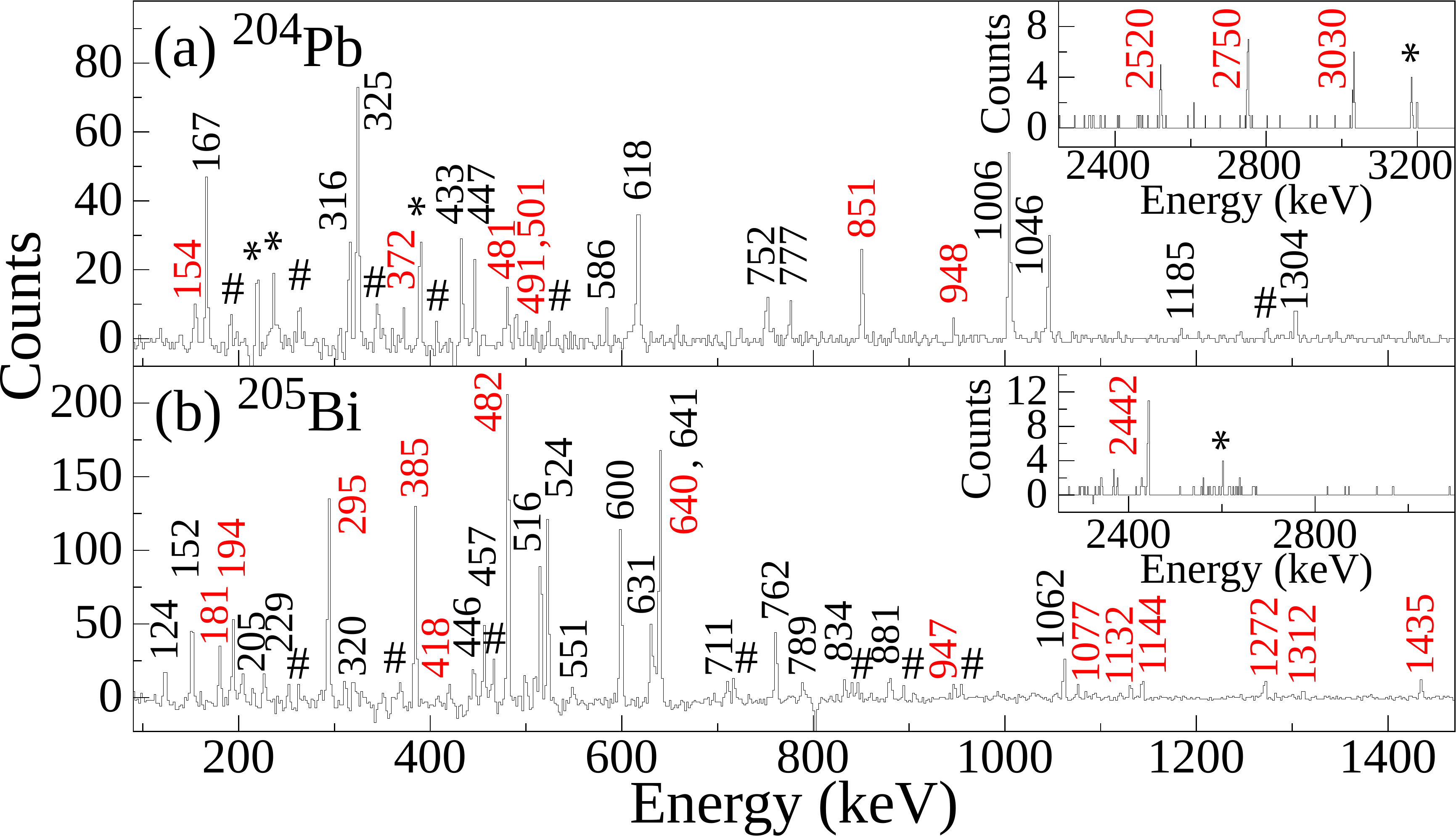}
\caption{Coincidence spectra illustrating $\gamma $ rays observed 
in the deexcitation of the {\it T}$_{1/2}$ = 0.22(2) ms and 8(2) ms isomers in 
(a) $^{204}$Pb and (b) $^{205}$Bi, respectively. Transitions with energy
$>$ 2 MeV are displayed in the insets. Asterisks and hash marks designate
unplaced and contaminant $\gamma $ rays, respectively. The ones in black font 
were established earlier \cite{Byrne1989_2,Linden1978}, while those in red have 
been identified from the present work.}
\label{fig:fig1}
\end{center}
\end{figure}

\begin{figure}
\begin{center}
\includegraphics[scale=0.4]{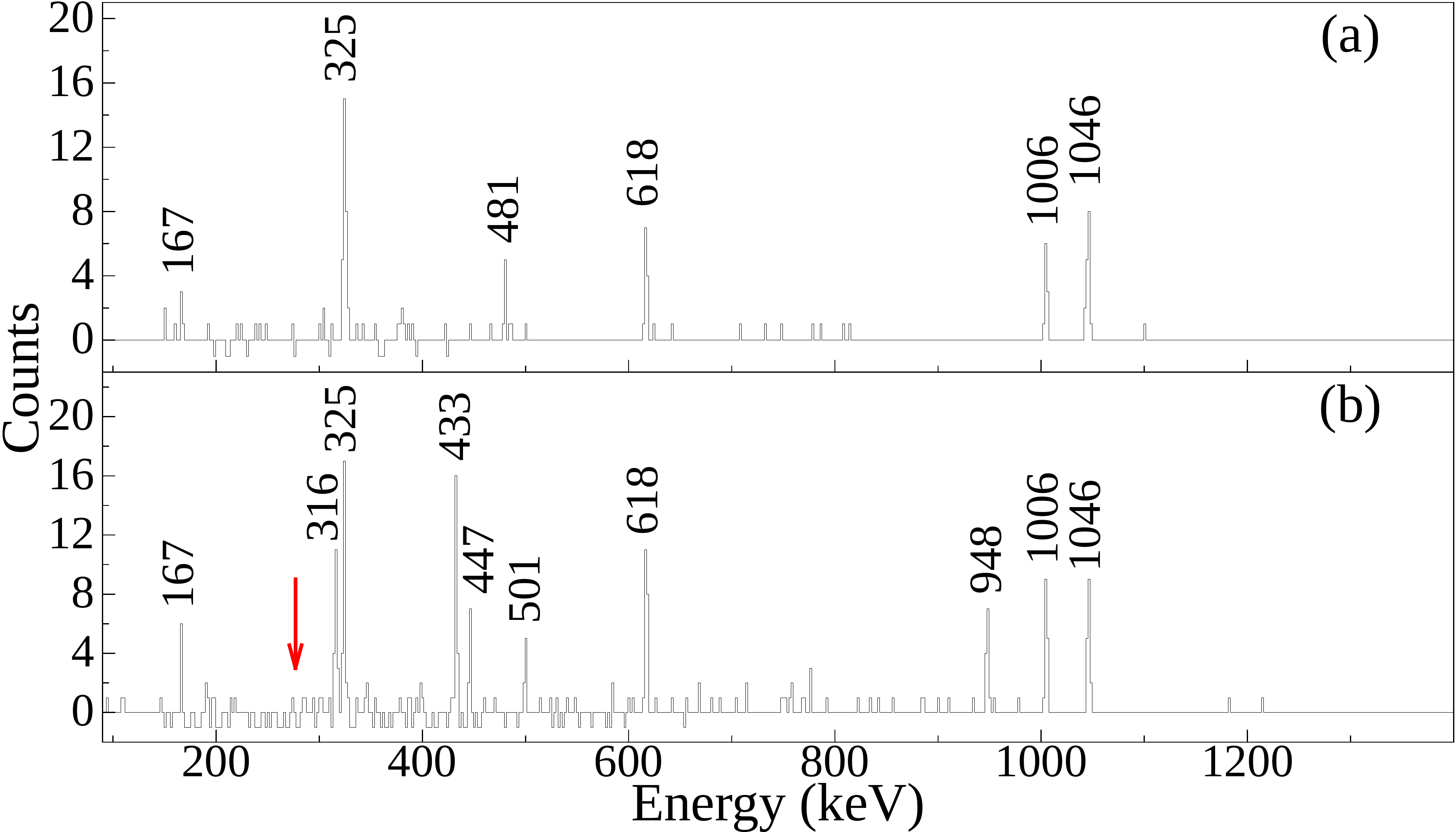}
\caption{Summed coincidence delayed spectra illustrating the decay of the
isomer in $^{204}$Pb: (a) the 2520-keV transition in coincidence with the previously
established 1006-325-618-1046 keV cascade (\cite{Linden1978}), (b) the
1304-keV $\gamma $ ray in coincidence with the 1006-325-618-1046-316-433 cascade
identified earlier. It is evident that, in the former instance, the 481-keV
$\gamma $ ray is the only newly identified one in addition to the gating
transition at 2520 keV. In the latter instance, only the 501- and 948-keV
transitions are newly established. The arrow indicates the absence, in the
delayed data, of the 277-keV, {\it E}$_{x}$ = 8126 keV to 7849 keV $\gamma $
ray observed in the prompt data from the previous work. This $\gamma $ ray
is clearly visible in the present prompt data, but not in the delayed spectrum.
The 481-, 501- and 948-keV peaks are evident only in the delayed spectra.
The present analysis, with the observed coincidence relationships, and the
energy sums for the above three parallel decay paths, unambiguously establish
the excitation energy of the isomer in $^{204}$Pb as 8349 keV.}
\label{fig:fig2}
\end{center}
\end{figure}

\clearpage

\begin{figure}
\begin{center}
\includegraphics[scale=0.3]{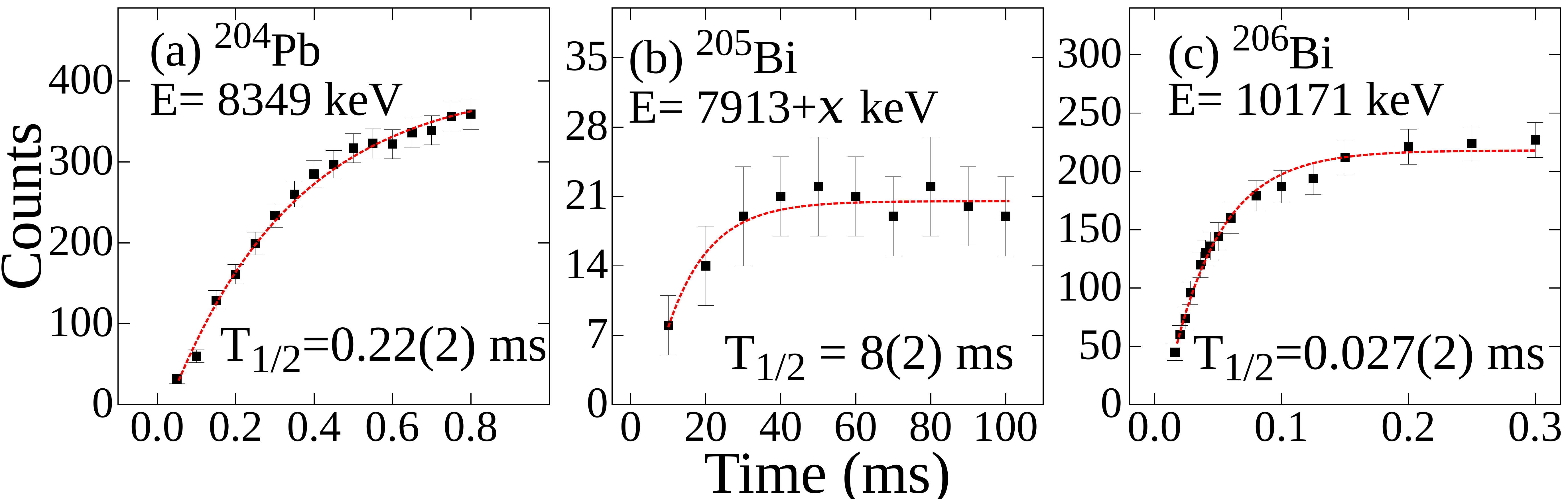}
\caption{Time distributions for the decay of isomers with excitation energy $\ge $ 8 MeV 
in: (a) $^{204}$Pb, (b) $^{205}$Bi, and (c) $^{206}$Bi. The integral counts, 
along with the half-lives inferred from the fits, are displayed.}
\label{fig:fig3}
\end{center}
\end{figure}

\begin{figure}
\includegraphics[scale=0.6]{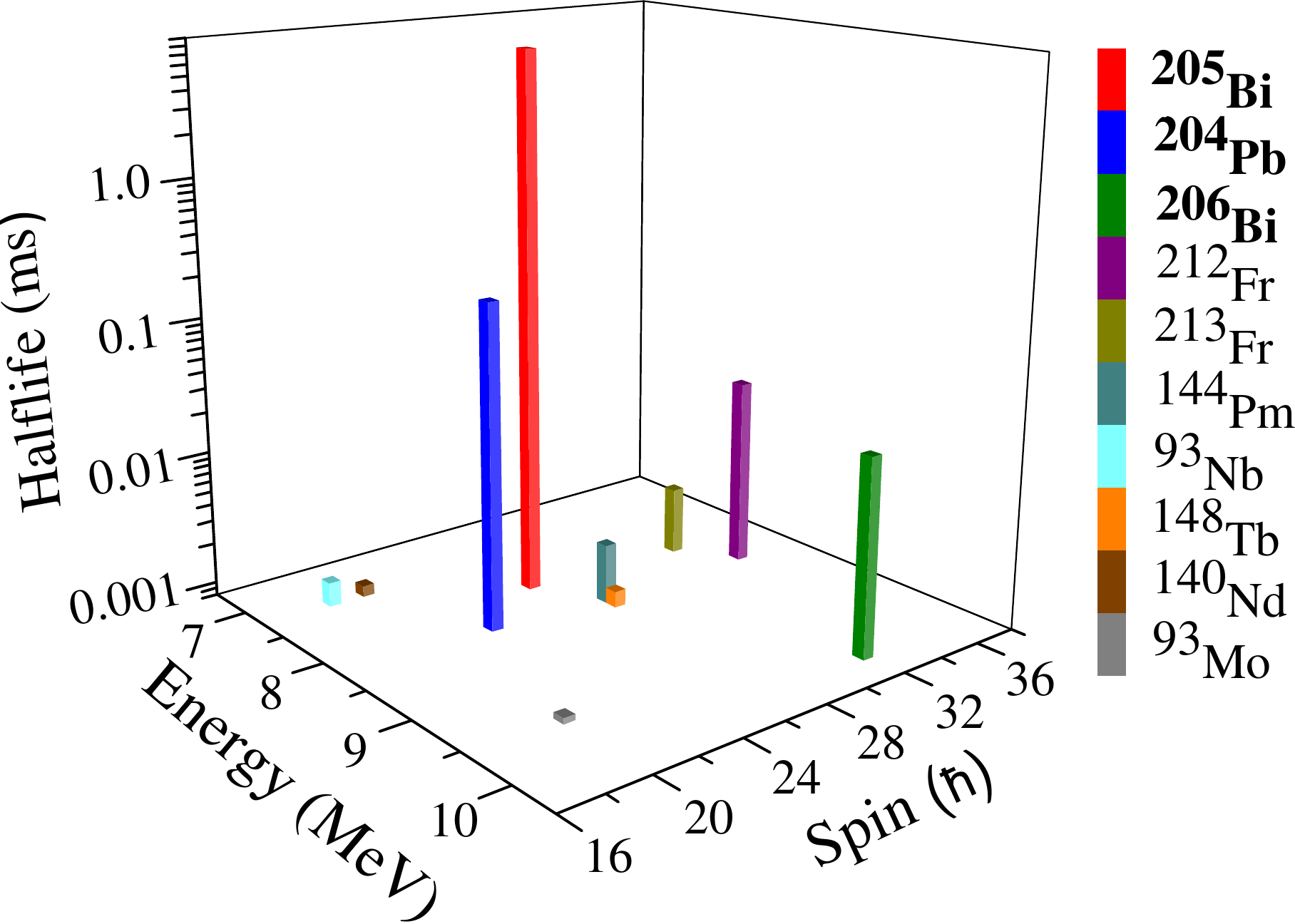}
\caption{Long-lived states ({\it T}$_{1/2}$ $>$ 1 $\mu $s) above an excitation energy of 7 MeV
in nuclides across the nuclear chart. Note that the half-lives are displayed on a logarithmic 
scale. The large difference between the half-lives of the isomers in $^{205}$Bi and $^{204}$Pb 
established from this work, and those in other nuclei, is evident.} 
\label{fig:fig4}
\end{figure}

\clearpage

\begin{figure}[htb]
\subfloat{%
    \includegraphics[angle=0,width=0.96\columnwidth]{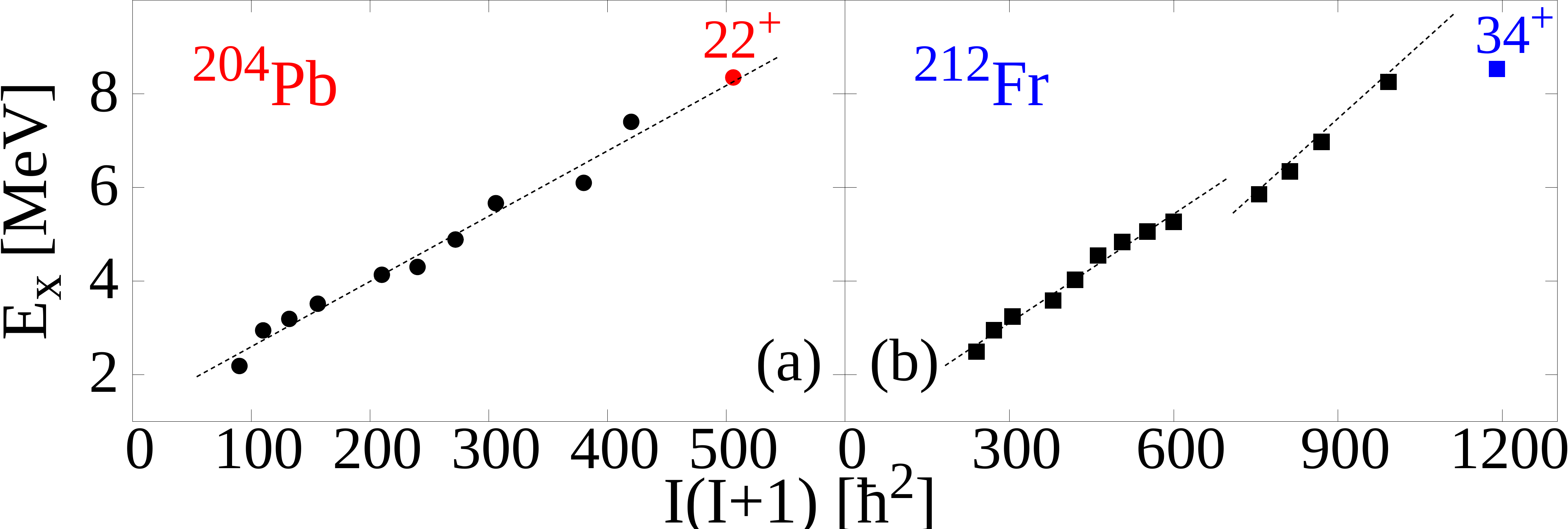}}
\newline
\subfloat{%
    \includegraphics[angle=0,width=0.96\columnwidth]{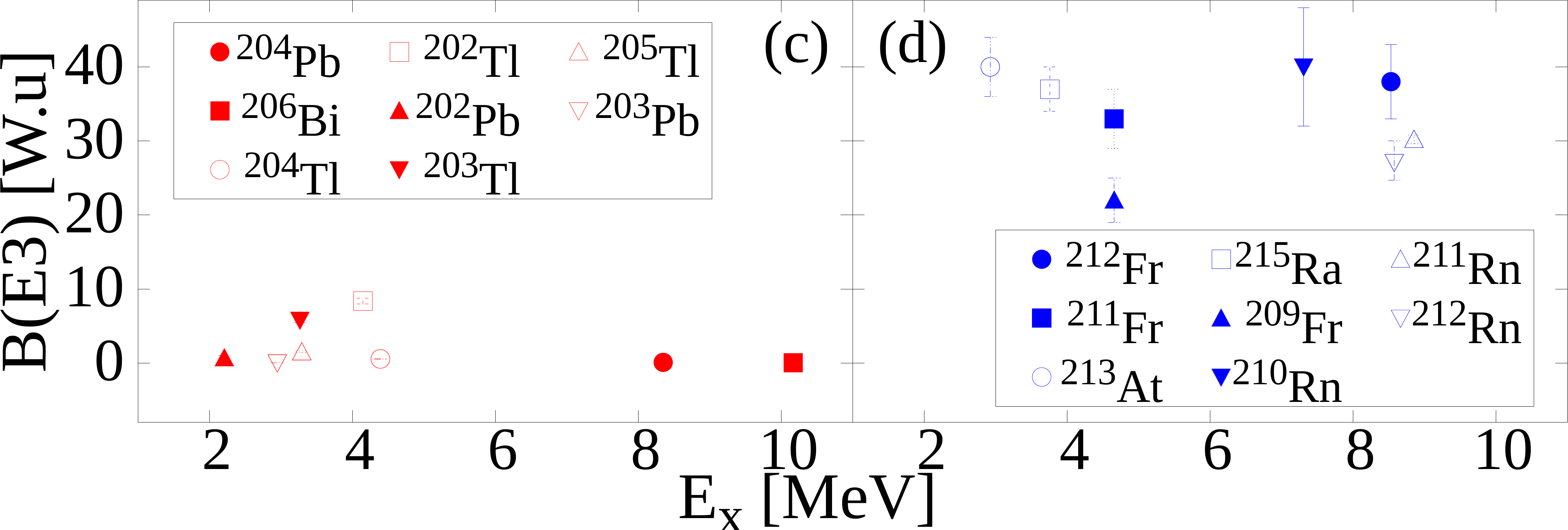}}
\caption{(a), (b) Locations of isomers in $^{204}$Pb and $^{212}$Fr
in the excitation energy ($E_x$) - angular momentum ($I$) plane. 
The dashed lines are intended to guide the eye. 
(c), (d) Reduced $E$3 transition probabilities [$B(E3)$] for isomer 
decays in the Tl-Pb-Bi and Rn-Fr-Ra region, respectively. 
The evidently different nature of these two types of isomers 
is discussed in the text.}
\label{fig:fig5}
\end{figure}

\end{document}